# Rotational instability in superlubric joints

Cangyu Qu[1,2], Songlin Shi[1,2], Quanshui Zheng[1,2,*]

[1] *Department of Engineering Mechanics, Tsinghua University, Beijing 100084, China*

[2] *Center for Nano and Micro Mechanics, Tsinghua University, Beijing 100084, China*

**Abstract**

Surface and interfacial energies play important roles in a number of instability phenomena in liquids and soft matters, but are rare to play a similar role in solids. Here we report a new type of mechanical instabilities that are controlled by surface and interfacial energies and are valid for a large class of materials, in particular two-dimensional layered materials. When sliding a top flake cleaved from a square microscale graphite mesa by using a probe acted on the flake through a point contact, we observed that the flake moved unrotationally for a certain distance before it suddenly transferred to a rotating-moving state. The theoretical analysis that agrees well with the experimental observation reveals that this mechanical instability is an interesting effect of the structural superlubricity (a state of nearly zero friction). Our further analysis shows that this type of instability holds generally for various sliding joints on different scales, as long as the friction is ultralow. Thus, the uncovered mechanism provides useful knowledge for manipulating and controlling all these sliding joints, and can guide design of future structural superlubricity based devices.

**Categories**:

   Mechanical instabilities are common phenomena that have important applications on one hand [1]–[4] and may result in serious calamities on the other hand [5]. Surface and interfacial energies play important roles in a number of instability phenomena in liquids or soft matters [6]–[9], but are rare to play a similar role in solids. For a micro/nano system, the surface-to-volume ratio is increased significantly, thus the effect of surface and interfacial energies is expected to enlarge. However, in sliding devices, this effect is usually overwhelmed by interfacial friction. This can be seen by comparing the frictional stress of an unlubricated contact, $\tau$, usually on the order of $\sim 600$ MPa (silica-silica [10]), with a typical value of surface energy, $\gamma$, on the order of $\sim 1$ J/m$^2$ (silicon(111) [11]). If the friction is assumed to be proportional to the contact area, and surface energy induced force proportional to a characteristic contact length, e.g. $L \sim 1$ μm, the frictional force $\tau L^2$ would be more than 2 orders larger than the surface energy induced force $\gamma L$. In order to see the effect of surface energy in these sliding joints, friction has to be ultralow. This is exactly the case in superlubric systems, where the incommensurability of the solid contact results in systematic cancellation of interfacial friction [12]. Since its theoretical promotion [13], [14], this fascinating phenomenon has drawn increasing interest, and has been demonstrated experimentally on various nano- to microscale systems [12], [15]. Especially, fabricated mesas of layered material demonstrate superlubricity through a self-retraction motion [16]. This layered-material-based (mostly graphite) sliding joint is suitable for future applications in micro devices for its compatibility with microfabrication process and the possibility of achieving superlubricity on micro or even larger scales [17].

In this Letter we report a new instability phenomenon observed on microscale graphite sliding joints. When sheared by a probe (either centric or eccentric), the top flake of this graphite mesa slides without rotation for a certain displacement, then suddenly transfers to a rotating-moving state. A simple model reveals that this rotational instability is controlled by the surface and interfacial energies of graphite, and is made possible by the nearly zero friction due to structural superlubricity. Our analysis, which is scaling-invariant and independent on the specific material, suggests that the same mechanism applies to not only graphite joints but generally sliding joints on different scales, as long as the friction is ultralow. This study provides useful knowledge for manipulating and controlling these joints, especially their relative angles, thus helping the future researches and applications of superlubricity.

Square shaped graphite mesas with side length 3-5 μm and height around 1 μm were fabricated from a highly oriented pyrolytic graphite (HOPG, Veeco, ZYB grade) using techniques developed previously [16]. Once fabricated, each mesa was sheared by a tungsten tip laterally, and cleaved into top and bottom flakes due to the layered structure of graphite, as shown in Fig.1(a). For around 60% of the total 60 sheared mesas, the interfacial friction is so small that the top flake retracts back to minimize its cleavage energy once the tip is released. This self-retraction phenomenon has been reported and utilized for the measurement of graphite surface/interface energy and investigations of superlubricity in previous publications [17]–[21].

During the manipulation of graphite mesas, a rotational instability has been repeatedly observed. As shown in Fig.1(b), when sheared by a nearly centric tip (quasi-statically with speed 10-100 nm/s), the top flake with its side length, $a$, slides without rotation for a displacement up to a critical value $\delta_{\text{cr}} \approx 0.9a$. And when the critical point is reached, the top flake suddenly rotates, demonstrating a rotational instability. On the other hand, when shearing using an eccentric tip with eccentricity, $e$, one might expect the top flake to rotate upon shearing due to the eccentric loading. However, our observation shows a different behavior. As shown in Fig.1(c), the top flake still slides without rotation for a considerable displacement, until similar rotational instability happens at a critical displacement, which is smaller than that in centric loading case. This can be better seen on the curve of the top flake rotation angle $\theta$ v.s. displacement $\delta$. As shown in Fig.1(d), $\theta$ remains nearly zero before a 0-order discontinuity in the curve corresponding to the rotational instability is found. Finally, it is noted that such sudden rotation usually locks the interface into a high-frictional commensurate state, and further shearing leads to the cleavage of a new layer [17].

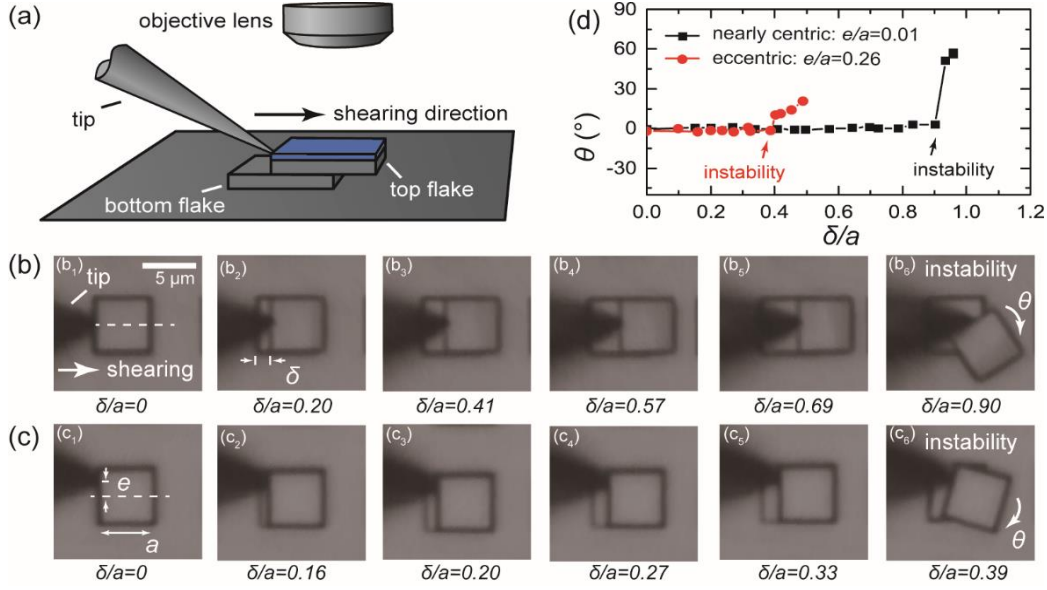

Figure 1. Experimental setup and the rotational instability. (a) The microscale graphite mesa is sheared by a tungsten tip laterally and observed by an optical microscope. (b)(c) Images from top view. Scale bar 5 μm. (b$_1$)-(b$_5$) When sheared by a nearly centric tip ($e/a \approx 0.01$), the top flake slides without rotation for a displacement up to $\delta_{cr}/a \approx 0.90$. (b$_6$) At the critical point $\delta_{cr}$, an instability characterized by a sudden rotation of the top flake is observed. (c$_1$)-(c$_5$) When sheared by an eccentric tip ($e/a \approx 0.26$), the top flake still slides without rotation for a displacement up to $\delta_{cr}/a \approx 0.39$. (c$_6$) At the critical point $\delta_{cr}$, similar rotational instability is observed. (d) Top flake rotation angle $\theta$ v.s. shearing displacement $\delta$ exacted from (b) and (c).

The experimental observations are rationalized by theoretical modelling. The system is modeled as a fixed bottom flake and a rigid but movable top flake with two degrees of freedom, i.e. the displacement along with the tip $\delta$, and the rotation around the tip contact point $\theta$, as shown in Figs. 2(a)(b). For an arbitrary state $(\delta, \theta)$, the total potential energy of the system is $U = 2\gamma_{sa}(A - A_{ss}) + \sigma A_{ss}$, where $\gamma_{sa}$ is the surface energy of graphite, $\sigma$ is the interfacial energy of the incommensurate graphite-graphite interface, i.e. the overlapping area, $A_{ss}$, between the top and bottom flakes during sliding, and $A = a^2$, as shown in Fig. 2(b). By noting that $\Gamma = 2\gamma_{sa} - \sigma$ which is the cleavage energy of graphite, we have

$$U(\delta, \theta) = 2\gamma_{sa}a^2 - \Gamma A_{ss}(\delta, \theta). \tag{1}$$

In a general case, $\sigma$ should depend on the misfit angle of the incommensurate interface, thus is $\theta$-dependent. However, such an angular dependence is typically small. Experimental measurements show that $\Gamma$ is nearly constant in the wide range of misfit angles accessible in experiments between two commensurate angles (0 and 60°) [18]. Thus we will approximate $\sigma$ (and thus $\Gamma$) to be constant in our model. In this regard, by defining the normalized total excess potential $U^* = (U - 2\gamma_{sa}a^2)/(\Gamma a^2) = -A_{ss}/a^2$, the calculation of $U^*$ comes down to the calculation of the overlapping area $A_{ss}$, which is conducted either analytically or with numerical algorithms [22]. Furthermore, the shearing is quasi-static, and the kinetic energy is neglected, as detailed in Supplemental Material.

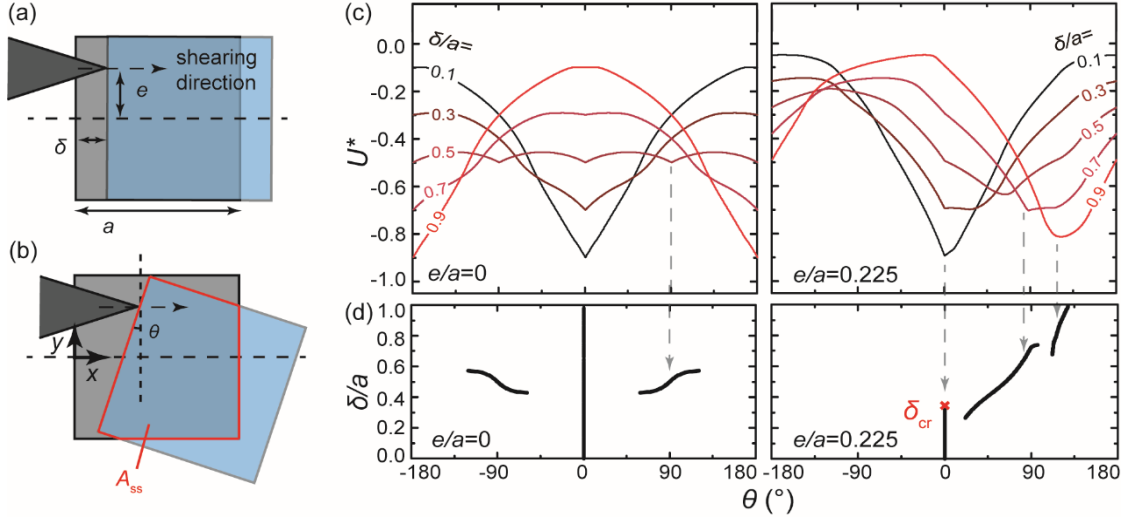

Figure.2 Theoretical calculations. (a) The top flake (side length $a$) is sheared by the tip (with an eccentricity $e$) in the $x$-direction by displacement $\delta$. (b) The rotation of the top flake around the tip contact point is denoted as $\theta$. (c) Calculated normalized total potentials ($U^*$) as functions of $\theta$ for centric loading ($e/a = 0$, left) and eccentric loading ($e/a = 0.225$, right), respectively. In the former case an energy minimum is always found at $\theta = 0$. Such a state is always stable/metastable for $\delta/a$ ranging from 0 to 1. In the latter case an energy minimum is found at $\theta = 0$ for $\delta < \delta_{cr} \approx 0.329a$. When $\delta_{cr}$ is reached, such an energy minimum vanishes, leading to the sudden rotation. (d) Branches of stable/metastable points other than $\theta = 0$ are also found as indicated by the black lines that correspond to the left- and right-hand sides of (c) in $(\delta, \theta)$ space where local energy minimum is met.

Figure 2(c) shows the calculated $U^*$ as functions of $\theta$ for $\delta/a$ ranging from 0 to 1, in the case of a perfectly centric tip loading ($e/a = 0$, left) and eccentric tip ($e/a = 0.225$, right), respectively. When the loading is centric, $\theta = 0$ is always a local energy minimum for the system regardless of $\delta$, indicating that un-rotated sliding is always a stable/metastable state. Considering the fact that at larger $\delta$, the energy barrier preventing the rotation becomes smaller, experimental perturbations might lead to premature rotation, thus the experimentally observed $\delta_{cr}/a \approx 0.9$ is slightly smaller than the theoretical prediction ($\delta_{cr} = a$). Meanwhile, two symmetric branches of local energy minima at $\theta$ around 45 ° to 135 °(and -45 ° to -135 °) that only exist for $\delta/a$ around 0.4 to 0.6 are also found. However, these two branches were not experimentally observed, since in most cases, a rotation out of $\theta = 0$ locks the interface into high frictional commensurate state and prevents further sliding of this interface [17].

On the other hand, different behavior is found for eccentric loading. In the case of $e/a = 0.225$, an energy minimum is still found at $\theta = 0$, however, only for $\delta < \delta_{cr} \approx 0.329a$. At $\delta_{cr}$, such an energy minimum vanishes, leading to the sudden rotation of the top flake. Meanwhile, two more branches of energy local minima are found at larger $\delta$, and even a second rotational instability is theoretically predicted at $\delta/a$ around 0.7. However, due to the same reason of lock-in after rotation, these branches were not observed experimentally.

The above described results are representative based on our calculation for $e/a$ ranging from 0 to 0.5. Generally, we define critical displacement $\delta_{\text{cr}}$ as the displacement where $\theta = 0$ is no longer a local energy minimum for the system. Analytical calculations, as detailed in Supplemental Material, yield the dependence of $\delta_{\text{cr}}$ on the tip eccentricity as:

$$\delta_{\text{cr}} = a - \sqrt{2ae} \tag{2}$$

This theoretical prediction is confirmed by experiments on 26 individual mesas, sheared by a tip with different eccentricities, until the rotational instability happens, as shown in Fig. 3. The results are consistent within experimental errors. Most data show smaller $\delta_{\text{cr}}$ than predicted. Among possible reasons, we find that a 2.5° misalignment of the shearing direction and mesa edge will result in a 20~30% lower $\delta_{\text{cr}}$, as detailed in Supplemental Material; vibration of tip that is inevitable will obviously also lead to premature rotation, yielding a lower $\delta_{\text{cr}}$. On the other hand, a few data show larger $\delta_{\text{cr}}$ than predicted, which is possibly due to delayed rotation resulted from the interfacial friction that act as a resistance torque against rotation.

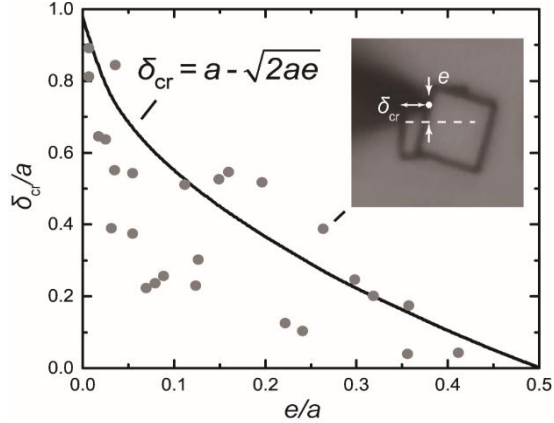

Figure 3. Dependence of $\delta_{\text{cr}}$ on $e$. The black curve indicates the theoretical prediction. The experimental results on 26 mesas, indicated by gray dots, are consistent with the theoretical prediction within experimental errors. Inset shows a typical snapshot of the system when the rotational instability happens.

In a more generalized case, the tip makes contact with the top flake not necessarily from the side, but at an arbitrary point $(e', e)$ on the surface of the top flake, as illustrated in the Inset of Fig.4(a), with $0 < e'/a < 1$; $-0.5 < e/a < 0.5$. This is the typical loading condition when shearing the mesa using an atomic force microscope (AFM) tip [21]. In this case, similar calculations, as detailed in Supplemental Material, lead to the dependence of $\delta_{\text{cr}}$ on $(e', e)$ as

$$\delta_{\text{cr}} = \begin{cases} a - e' + \sqrt{e'^2 - 2ae}, & \text{for } 2ae - e'^2 \leq 0 \\ \max\left(0, a - e' - \sqrt{2ae - e'^2}\right), & \text{for } 2ae - e'^2 > 0 \end{cases} \tag{3}$$

The $\delta_{\text{cr}}$-contour on the $(e', e)$ plane is plotted in Fig.4(a). For $e = 0$, $\delta_{\text{cr}}$ equals $a$ regardless of $e'$, indicating that the sliding remains un-rotated until the end of shearing.

Meanwhile, at other positions where $\delta_{cr}/a$ is between 0 and 1, the rotational instability happens at a certain point given by Eq. (3). Notably, $\delta_{cr} = 0$ is found in certain regions on the $(e', e)$ plane, as indicated as the light gray domains in Fig. 4(a). In other words, the top flake would rotate immediately upon shearing when the tip contact point is located in these domains. This prediction agrees with the experimental observations as shown in Figs 4(b) and 4(c). The same mesa is sheared by an AFM (NT-MDT, NTEGRA Prima) tip contacting the top flake with constant normal load of around 10 μN. When shearing from point $(e'/a, e/a)=(0.51, 0.06)$ (Fig.4(b)), the top flake moves without rotation; while when shearing from point $(e'/a, e/a)=(0.51, 0.35)$ (Fig.4(c)), the top flake rotates immediately upon shearing. In the above discussion, the tip-mesa contact is assumed to be frictionless. While in experiments, a frictional torque at the contact might provide additional resistance against the rotation.

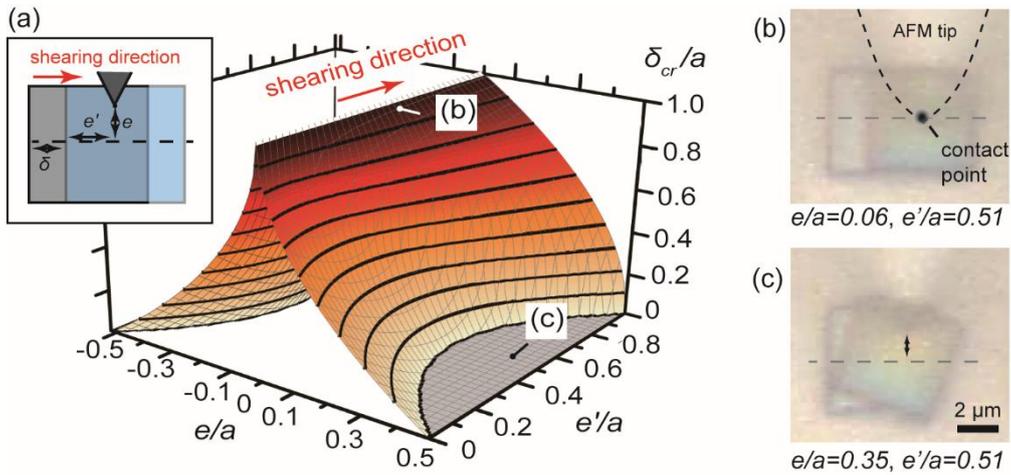

Figure 4. Loading from a tip contact point $(e', e)$. (a) $\delta_{cr}$-contour on the $(e', e)$ plane. For $e = 0$, $\delta_{cr}/a$ equals 1 regardless of $e'$, indicating that the sliding remains un-rotated until the end. At other loading positions where $\delta_{cr}/a$ is between 0 and 1, the sliding remains un-rotated until rotational instability happens. And notably, $\delta_{cr} = 0$ is found in two symmetric regions in the phase space, as indicated by the light gray domains. This indicates that the top flake rotates immediately when sheared at these positions. Inset shows the sketch of the system. (b)(c) experimental images of shearing a mesa from two different contact points, i.e. $(e'/a, e/a)=(0.51, 0.06)$ (b), and $(e'/a, e/a)=(0.51, 0.35)$ (c). The top flake moves without rotation in (b); while rotates immediately upon shearing in (c). The results are consistent with the theoretical predictions.

The rotational instability in the graphite sliding joints is directly dependent on the geometry of the system, thus we further investigated the instability of rectangular mesas with different aspect ratio, and circular mesas. Similar instability behaviors were found and discussed in Supplemental Material. We emphasize that this type of instability is a common phenomenon in these systems. In fact, it can be seen from Eq. (1) that the $\delta$- and $\theta$-dependence of the system potential energy comes from $A_{ss}$, which is purely determined by geometry. Thus the behaviors discussed above and the values of critical points are independent on the specific material, which only influence the overall magnitude of system potential through the material constant $\Gamma$. Furthermore, it is noted that all the parameters in the discussion is normalized against the characteristic length of the system, here, the side length of the mesa $a$. Hence the major conclusions of the analysis, e.g. the critical displacement $\delta_{\text{cr}}/a$, are dimensionless quantities independent on $a$. In this regard, the results are scaling-invariant, unless new physics has to be added into the model, e.g. when structural superlubricity fails on larger scale and interfacial friction has to be considered.

One of the main challenges in superlubric devices is to control the sliding direction and constrain relative rotation without introducing extra mechanical contacts (where friction is difficult to avoid)[23], [24]. The mechanism behind the rotational instability provides an elegant idea of effectively constraining relative rotation through interfacial energy. This idea, for its simplicity, could be extended to any geometry of interest. As a demonstration, we add a case study on a superlubric slider sliding on a guide rail, and show that it is expected to be rotation-proof, as detailed in Supplemental Material. In this regard, the reported instability phenomenon and its mechanism provide guidance for design of a class of general superlubric devices.

In conclusion, we observed a novel rotational instability in microscale graphite sliding joints. The comparison between experimental observation and theoretical modelling reveals that such instability is controlled by the surface and interfacial energies of graphite. The result is independent on the specific material and is fully scalable, thus is not limited to graphite, but also applicable to other superlubric sliding systems. Our results provide useful knowledge for manipulating and controlling these sliding joints, especially, constraining relative rotation through interfacial energy and design of the device geometry.

**Acknowledgement**

We thank very much Wen Wang and Diwei Shi for the valuable discussions. This work is supported by the National Key Basic Research Program of China (Grant No. 2013CB934200), the National Natural Science Foundation of China No.11572173, No. 11772168 and No. 11890670, No.11632009，No.11372153，No.11172149, the Cyrus Tang Foundation through Grant No. 202003.